\documentclass[twocolumn,amsmath,showpacs,prl,superscriptaddress]{revtex4}

\newcommand{ \be }{\begin{equation}}
\newcommand{ \ee }{\end{equation}}
\newcommand{ \bea }{\begin{eqnarray}}
\newcommand{ \eea }{\end{eqnarray}}

\usepackage{graphicx}
\usepackage{dcolumn} 
\usepackage{bm} 
\usepackage[dvips]{color}

\def\ARNPS{ Ann. Rev. Nucl. Part. Sci.}
\def\PLB{{ Phys. Lett.}  B}
\def\NPA{{ Nucl. Phys.}  A}

\def\PRL{ Phys. Rev. Lett.}
\def\PRC{{ Phys. Rev.} C}

\def\PR{ Phys. Rep.}

\def\Journal#1#2#3#4{{#1} {\bf #2}, #3 (#4)}

\begin{document}

\title{Azimuthally Sensitive Hanbury Brown-Twiss Interferometry in Au+Au Collisions at $\sqrt{s_{NN}}=200$ GeV}

\affiliation{Argonne National Laboratory, Argonne, Illinois 60439}
\affiliation{Brookhaven National Laboratory, Upton, New York 11973}
\affiliation{University of Birmingham, Birmingham, United Kingdom}
\affiliation{University of California, Berkeley, California 94720}
\affiliation{University of California, Davis, California 95616}
\affiliation{University of California, Los Angeles, California 90095}
\affiliation{California Institute of Technology, Pasadena, California 91125}
\affiliation{Carnegie Mellon University, Pittsburgh, Pennsylvania 15213}
\affiliation{Creighton University, Omaha, Nebraska 68178}
\affiliation{Nuclear Physics Institute AS CR, \v{R}e\v{z}/Prague, Czech Republic}
\affiliation{Laboratory for High Energy (JINR), Dubna, Russia}
\affiliation{Particle Physics Laboratory (JINR), Dubna, Russia}
\affiliation{University of Frankfurt, Frankfurt, Germany}
\affiliation{Indian Institute of Technology, Mumbai, India}
\affiliation{Indiana University, Bloomington, Indiana 47408}
\affiliation{Institute  of Physics, Bhubaneswar 751005, India}
\affiliation{Institut de Recherches Subatomiques, Strasbourg, France}
\affiliation{University of Jammu, Jammu 180001, India}
\affiliation{Kent State University, Kent, Ohio 44242}
\affiliation{Lawrence Berkeley National Laboratory, Berkeley, California 94720}\affiliation{Max-Planck-Institut f\"ur Physik, Munich, Germany}
\affiliation{Michigan State University, East Lansing, Michigan 48824}
\affiliation{Moscow Engineering Physics Institute, Moscow Russia}
\affiliation{City College of New York, New York City, New York 10031}
\affiliation{NIKHEF, Amsterdam, The Netherlands}
\affiliation{Ohio State University, Columbus, Ohio 43210}
\affiliation{Panjab University, Chandigarh 160014, India}
\affiliation{Pennsylvania State University, University Park, Pennsylvania 16802}
\affiliation{Institute of High Energy Physics, Protvino, Russia}
\affiliation{Purdue University, West Lafayette, Indiana 47907}
\affiliation{University of Rajasthan, Jaipur 302004, India}
\affiliation{Rice University, Houston, Texas 77251}
\affiliation{Universidade de Sao Paulo, Sao Paulo, Brazil}
\affiliation{University of Science \& Technology of China, Anhui 230027, China}
\affiliation{Shanghai Institute of Nuclear Research, Shanghai 201800, P.R. China}
\affiliation{SUBATECH, Nantes, France}
\affiliation{Texas A\&M University, College Station, Texas 77843}
\affiliation{University of Texas, Austin, Texas 78712}
\affiliation{Valparaiso University, Valparaiso, Indiana 46383}
\affiliation{Variable Energy Cyclotron Centre, Kolkata 700064, India}
\affiliation{Warsaw University of Technology, Warsaw, Poland}
\affiliation{University of Washington, Seattle, Washington 98195}
\affiliation{Wayne State University, Detroit, Michigan 48201}
\affiliation{Institute of Particle Physics, CCNU (HZNU), Wuhan, 430079 China}
\affiliation{Yale University, New Haven, Connecticut 06520}
\affiliation{University of Zagreb, Zagreb, HR-10002, Croatia}
\author{J.~Adams}\affiliation{University of Birmingham, Birmingham, United Kingdom}
\author{C.~Adler}\affiliation{University of Frankfurt, Frankfurt, Germany}
\author{M.M.~Aggarwal}\affiliation{Panjab University, Chandigarh 160014, India}
\author{Z.~Ahammed}\affiliation{Variable Energy Cyclotron Centre, Kolkata 700064, India}
\author{J.~Amonett}\affiliation{Kent State University, Kent, Ohio 44242}
\author{B.D.~Anderson}\affiliation{Kent State University, Kent, Ohio 44242}
\author{D.~Arkhipkin}\affiliation{Particle Physics Laboratory (JINR), Dubna, Russia}
\author{G.S.~Averichev}\affiliation{Laboratory for High Energy (JINR), Dubna, Russia}
\author{S.K.~Badyal}\affiliation{University of Jammu, Jammu 180001, India}
\author{J.~Balewski}\affiliation{Indiana University, Bloomington, Indiana 47408}
% ...

\collaboration{STAR Collaboration}\noaffiliation

\date{\today}

\begin{abstract}
We present the results of a systematic study of the shape of the pion distribution in coordinate space at freeze-out in Au+Au collisions at RHIC using two-pion Hanbury Brown-Twiss (HBT) interferometry. Oscillations of the extracted HBT radii vs. emission angle indicate sources elongated 
perpendicular to the reaction plane.  The results indicate 
that the pressure and expansion time of the collision system are not sufficient to completely quench its initial shape.
\end{abstract}


\maketitle

Relativistic heavy ion collisions are believed to reach sufficiently high energy densities and temperatures for the possible
formation of a quark-gluon plasma (QGP) ~\cite{qgp3}.  Hanbury Brown-Twiss (HBT) interferometry~\cite{heinzJacak} 
of two particle Bose-Einstein correlations 
directly accesses the space-time structure of the emitting source formed in these collisions,
providing crucial probes of the system dynamics.  At the Relativistic Heavy 
Ion Collider (RHIC), identical-pion HBT studies in Au+Au collisions at $\sqrt{s_{NN}}=130$ GeV~\cite{starHbtPrl,phenixHbtPrl} 
yielded an apparent source size consistent with measurements at lower energies, in contrast 
to predictions of larger sources based on QGP formation~\cite{RischkeGyulassy}.  In addition, hydrodynamical models, 
successful at RHIC in describing transverse momentum spectra and elliptic 
flow~\cite{starV2Prl}, have failed to reproduce the small HBT radii~\cite{heinzKolbWW02}.  This so-called
``HBT puzzle"~\cite{heinzPanic02,dumitru02} might arise because the system's lifetime is
shorter than predicted by models.

In non-central collisions, azimuthally-sensitive HBT measurements performed relative to the reaction plane
provide a measure of the source shape at freeze-out~\cite{wiedemannPR,e895Plb,voloshinCleland}.
In such collisions, the almond-shaped collision geometry generates greater transverse pressure gradients in
the reaction plane than perpendicular to it.  This leads to
stronger {\it in-plane} expansion (elliptic flow)~\cite{KSH00,voloshinQM,starV2Prl,olli} which diminishes the initial {\it out-of-plane}
spatial anisotropy.  Therefore the freeze-out source shape should be sensitive to the evolution of the pressure
gradients and the system lifetime; a long-lived system would be less {\it out-of-plane} extended and perhaps {\it in-plane} extended.
Hydrodynamic calculations ~\cite{kolbQM02} predict a strong sensitivity of the HBT
parameters to the early conditions in the collision system and show that, while 
the system may still be out-of-plane extended after 
hydrodynamic evolution, a subsequent rescattering phase \cite{teaney} tends to make the final source in-plane.  Knowledge of the freeze-out source shape might 
discriminate among scenarios of the system's evolution.

In this Letter, we present results of a systematic study of azimuthally-sensitive HBT in Au+Au collisions 
at $\sqrt{s_{NN}}=200$ GeV.  
These 
results allow for first studies of the relationship between the initial and final eccentricities of 
the system. 

The measurements were made using the STAR detector~\cite{starNim} at RHIC.  Particle trajectories and momenta were reconstructed using a Time Projection 
Chamber (TPC) with full azimuthal coverage, located inside a 0.5 Tesla solenoidal magnet. Au+Au events with primary 
vertices $\leq 25$ cm longitudinally of the TPC center were placed into centrality classes 
following Ref.~\cite{starHadron}.  A high-multiplicity triggered dataset of ~500k events 
was used for the most central bin (0--5\% total cross section), and a minimum bias dataset of 1.6 million events 
was used for all other centrality classes (5--10\%, 10--20\%, 20--30\% and 30--80\%).  
The second-order event plane angle $\Psi_2$~\cite{poskanzer} for each
event was determined from the weighted sum of primary charged-particle
transverse momenta ~\cite{STARfirstv2}.
Within a resolution which we determine from the random subevent
method~\cite{poskanzer}, $\Psi_2 \approx \Psi_{\rm rp}$ (true reaction plane angle)
or~$\Psi_2 \approx \Psi_{\rm rp}+\pi$;
{\it i.e.} the direction of the impact parameter vector is determined up to
a sign~\cite{poskanzer,STARv1v4}.

Pion candidates, selected according to their specific energy loss ($dE/dx$) in the TPC in the rapidity range $|y|<0.5$, were required to pass within 3 cm 
of the primary vertex and contain $> 15$ (out of 45) TPC space points in the reconstructed trajectory.  Pion pairs were subjected to two requirements.  To account for reconstructing a single 
particle trajectory as two tracks, a topological cut is applied in which a minimum fraction of TPC pad 
layers must show distinct hits for both tracks.  To reduce the effect of merging two particle trajectories into a single 
reconstructed track, an additional topological cut requires that the number of {\it merged} TPC hits falls 
below a maximum fraction.  The latter cut leads to a systematic error that depends on the event multiplicity 
and the transverse momentum of the tracks \cite{starHbtPrl}. 

Pairs of like-sign pions were placed into bins of 
$\Phi^\prime \equiv \phi_{\rm{pair}} - \Psi_2$, where $\phi_{\rm{pair}}$ is the
azimuthal angle of the pair momentum [${\mathbf k}= \frac{1}{2} ({\mathbf p_1}+{\mathbf p_2})$].
Because we use the 2$^{\rm nd}$-order reaction plane, $\Phi^\prime$ is only defined in the range $(0,\pi)$.
For each bin, a three-dimensional 
correlation function is constructed in the Pratt-Bertsch ``out-side-long" decomposition~\cite{pb} 
of the relative pair momentum ${\mathbf q}$.  The numerator of the correlation function contains 
pairs of pions from the same event, and the denominator contains pairs of pions from different events
which have similar primary vertex position, reaction plane orientation, multiplicity, and magnetic field 
orientation.
$\pi^-$ pairs and $\pi^+$ pairs were mixed separately due to 
charge-dependent acceptances but are combined to increase statistics; separate $\pi^+$ and $\pi^-$ analyses showed no significant differences.

Finite reaction plane resolution and finite width of the $\Phi^\prime$ bins reduce the measured oscillation amplitudes of HBT radii vs.~$\Phi^\prime$.  A model-independent correction procedure \cite{HHLW}, applied to each ${\mathbf q}$-bin in the numerator and 
denominator of each correlation function, accounts for these effects and increases the 
amplitudes of the HBT radii vs.~$\Phi$ ($\Phi \equiv \phi_{\rm{pair}} - \Psi_{\rm{rp}}$). The increase is roughly inversely proportional to the  measured \cite{poskanzer,STARfirstv2} reaction plane resolution, {\it i.e.}~the amplitudes increase $\sim$10--30\%.  All data were corrected using this procedure.  Also, auto-correlation contributions to $\Phi$ were tested by selecting distinct sets of particles for event plane determination and HBT analysis, with no observed effect.

In addition, correlations due to final-state Coulomb repulsion must be accounted for, in order to 
isolate the Bose-Einstein correlations of interest.  Traditionally this was accomplished by applying correction
weights (determined by calculating the Coulomb correlation function $K({\mathbf q})$ for a spherical Gaussian source \cite{starHbtPrl}) to all pairs in the denominator.
Recently, the CERES collaboration~\cite{ceresNpa} noted that this approach over-corrects
for the Coulomb effect and advocated an improved procedure \cite{bowlersinyukov} which applies the Coulomb weight
only to the fraction of pairs that participate in the Bose-Einstein correlation.  We adopt this approach, and fit each experimental correlation function to the form:
\begin{equation}\label{eq:bowler}
C({\mathbf q},\Phi) = N \cdot \bigl[ (1-\lambda)\cdot 1 + \lambda \cdot K({\mathbf q}) \bigl(1 + G({\mathbf q},\Phi) \bigr)\bigr] ,
\end{equation}
where the $(1-\lambda)$  and $\lambda$ terms account for the non-participating and participating 
fractions of pairs, respectively, $N$ is a normalization parameter, and  $G({\mathbf q},\Phi)$ is 
the Gaussian correlation model~\cite{pb}:
\begin{equation}\label{eq:prattBertsch}
G({\mathbf q},\Phi) = e^{ - q_o^2 R_o^2(\Phi) - q_s^2 R_s^2(\Phi) - q_l^2 R_l^2(\Phi) - q_o q_s R_{os}^2(\Phi)} .
\end{equation}
$R^2_i$ are the squared HBT radii,
where the $l$,$s$,$o$ subscripts indicate the long (parallel to beam), side (perpendicular to beam and 
total pair momentum) and out (perpendicular to $q_l$ and $q_s$) decomposition of ${\mathbf q}$ with an additional cross term \cite{generalForm}.  Fitting with Eq.~\ref{eq:bowler} caused $R_o$ to increase ~10--20\% compared to Coulomb-correcting all pairs,  
 while $R_s$ and $R_l$, respectively, are consistent within errors.

\begin{figure} 
	{\includegraphics*[width=3.3in]{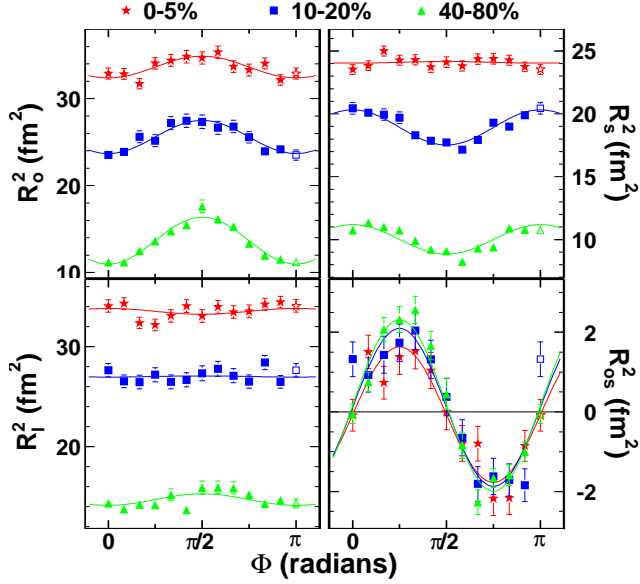}}
  \caption{(color online) Squared HBT radii using Eq.~\ref{eq:bowler} relative to the reaction plane angle 
for three centrality classes.  The solid lines show allowed~\cite{HHLW} fits to the individual 
oscillations.}\label{figure1}
\end{figure} 

Figure~\ref{figure1} shows the squared HBT radii, obtained using Eq.~\ref{eq:bowler}, as a function of $\Phi$ 
for three centrality classes.
All pairs 
with pair transverse momentum $0.15 \leq k_T \leq 0.6$ GeV/$c$ are included, and each centrality is divided into 
12 $\Phi$ bins of 15$^\circ$ width.
The data point at $\Phi=\pi$ is the reflected $\Phi=0$ value, and solid lines indicate Fourier expansions of the allowed oscillations~\cite{HHLW}:
\begin{equation}\label{fourier}
R^2_{\mu,n}(k_T) =
\begin{cases}
\langle R^2_\mu(k_T,\Phi) \cos(n\Phi) \rangle & (\mu = o, s, l) \\
\langle R^2_\mu(k_T,\Phi) \sin(n\Phi) \rangle & (\mu = os)
\end{cases}.
\end{equation}
As expected~\cite{starHbtPrl}, the 0$^{\rm{th}}$-order Fourier 
coefficient (FC) indicates larger apparent source sizes for more central 
collisions.  We verified that the 0$^{\rm{th}}$-order FC corresponds to the HBT radii from an 
azimuthally-integrated analysis.

Strong 2$^{\rm{nd}}$-order oscillations are observed for $R_o^2$, $R_s^2$ and $R_{os}^2$, and the signs 
of the oscillations are qualitatively self-consistent~\cite{wiedemannPR, HHLW}, though the amplitude for 
most-central events is small.  Similar oscillations were observed in a statistics-limited analysis of minimum-bias Au+Au collisions at $\sqrt{s_{NN}}=130$ GeV \cite{RandyThesis}.  These oscillations correspond to a pion source spatially extended perpendicular to the reaction 
plane, as discussed below.  The next terms (4$^{\rm th}$-order) in the Fourier expansions (Eq.~\ref{fourier}) are consistent with zero within statistical errors. 

\begin{figure} 
	{\includegraphics*[width=3.3in]{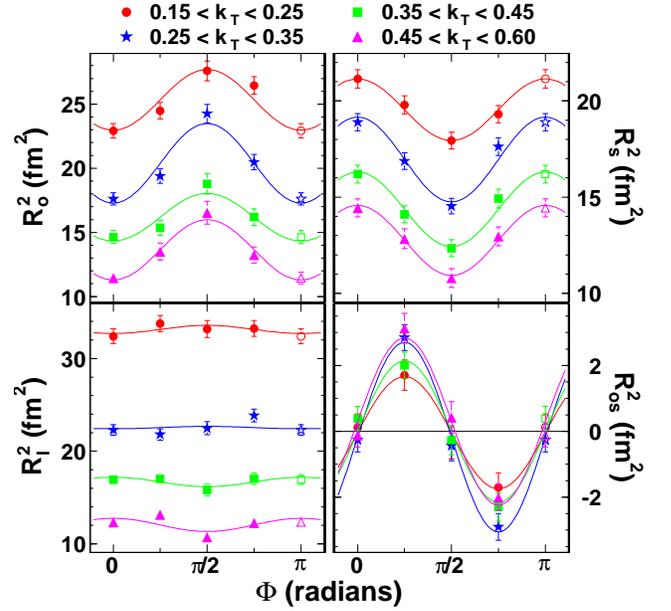}}
  \caption{(color online) Squared HBT radii relative to reaction plane angle for four $k_T$ (GeV/$c$) bins, 20--30\% centrality 
events. The solid lines show allowed~\cite{HHLW} fits to the individual oscillations.} \label{figure2}
\end{figure} 

The $k_T$-dependence of the oscillations of the HBT radii may contain
important information on the initial conditions
and equation of state of the system~\cite{heinzKolbPlb}.
Figure~\ref{figure2} shows the $\Phi$ dependence of HBT radii for mid-central (20--30\%) events for four 
$k_T$ bins.  Due to the additional division of pairs in $k_T$, only four bins 
in $\Phi$ are used.  
The 0$^{\rm{th}}$-order FC increases with decreasing $k_T$, which was observed for azimuthally-integrated 
HBT analyses at $\sqrt{s_{NN}}=130$ GeV~\cite{starHbtPrl} and attributed to pion emission from an expanding 
source. Strong out-of-plane oscillations are observed for all transverse radii in each $k_T$ bin.  

\begin{figure}[bt]
	{\includegraphics*[width=3.3in]{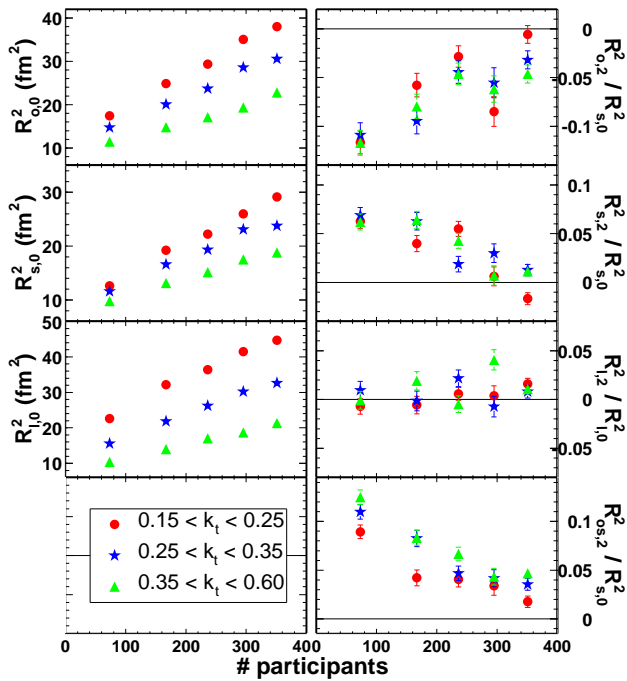}}
  \caption{(color online) Fourier coefficients of azimuthal oscillations of HBT radii vs.~number of participating nucleons, 
for three $k_T$ (GeV/$c$) bins.  Left panels: means ($0^{\rm{th}}$-order FC) of oscillations; right panels: relative 
amplitudes (see text for details).  Larger participant numbers correspond to more central collisions.}\label{figure3}
\end{figure} 

The full results are summarized in Figure~\ref{figure3}, which shows the centrality dependence of the Fourier
coefficients  for three ranges of $k_T$.  The number of participants for each centrality was determined using
a simple nuclear overlap model \cite{starHadron}.  Systematic variations of the HBT radii arise due to their sensitivity to the anti-merging cut threshold and uncertainty associated with
the Coulomb procedure \cite{starHbtPrl}.  The total variation is largest for $R_{o,0}^2$ ($\sim 10\%$).  The systematic variation on the relative amplitudes plotted in the right panels of Fig.~\ref{figure3} are negligible compared to statistical errors.  Also, all correlation functions composing Fig.~\ref{figure3} are corrected for momentum resolution following our prescription in Ref.~\cite{starHbtPrl}.

As in Figs.~\ref{figure1} and~\ref{figure2}, the 0$^{\rm{th}}$-order FCs 
(left panels) correspond to the squared HBT radii 
that would be obtained in a standard analysis.  $R_{o,0}^2$, $R_{s,0}^2$ and $R_{l,0}^2$ are all observed 
to decrease for more peripheral collisions. 
$R_o/R_s$, found in theoretical calculations to be
sensitive to the emission duration of the system \cite{RischkeGyulassy}, is observed to be 
$R_{o,0}/R_{s,0} =  1.15 \pm 0.01$ ($1.06 \pm 0.01$) for the lowest (highest) $k_T$ bin for 0--5\% most-central events.  
These values are consistent with that reported at $\sqrt{s_{NN}}=130$ GeV~\cite{starHbtPrl}
when the increase in $R_o$ due to the improved Coulomb correction (Eq.~\ref{eq:bowler}) 
is accounted for.  $R_o/R_s$ 
is still smaller than the predictions from hydrodynamical models, indicating the ``HBT puzzle" persists at 
$\sqrt{s_{NN}}=200$ GeV.  

Dynamical effects on the homogeneity region affect $R_{\mu,2}^2(k_T)$ as well as 
$R_{\mu,0}^2$~\cite{kolbQM02,fabriceMike}.  The {\it relative} amplitudes of the oscillations
offer a more robust measure of the spatial anisotropy and are less sensitive to dynamical effects \cite{fabriceMike}.  
Figure~\ref{figure3} shows (right panels) the 
relative amplitudes vs.~number of participants for three $k_T$ ranges, using the ratios $R_{\alpha,2}^2/R_{s,0}^2$ $(\alpha = o,s,os)$ and $R_{l,2}^2/R_{l,0}^2$.  
The relative amplitudes for all three transverse radii 
decrease in magnitude with increasing number of participants, and their weak $k_T$ dependence agrees qualitatively with hydrodynamic calculations \cite{kolbQM02}.  

To extract the shape of the pion source at freeze-out, a model-dependent approach is 
required.  In the presence of collective flow the HBT radii correspond 
to regions of homogeneity~\cite{Sinyukov_homogeneity} and do not reflect the entire source.  
The ``blast-wave" parametrization~\cite{starV2Prl,schnedermann,huovinen,fabriceMike}
of freeze-out, which incorporates
both spatial and dynamical anisotropies, has been used to describe various observables at $\sqrt{s_{NN}}=130$ GeV \cite{fabriceMike,starKpi}.  A recent blast-wave analysis \cite{fabriceMike} showed 
that the relative oscillation amplitudes ({\it e.g.}~shown in Fig.~3) are most sensitive to the spatial anisotropy. 
The source eccentricity $\bigl(\varepsilon\equiv ({R_{y}^2-R_{x}^2})/({R_{y}^2+R_{x}^2})\bigl)$ can be related to the relative amplitude of the HBT oscillations by $\varepsilon_{\rm final} \approx 2 {R_{s,2}^2}/{R_{s,0}^2}$ \cite{wiedemannPR,fabriceMike},
where $R_x$ ($R_y$) is the radius of the elliptical source in-plane (out-of-plane).  

The eccentricity of the initial almond-shaped overlap region was calculated from a Glauber model \cite{starHadron} using the {\it r.m.s.}~values for $R_y$ and $R_x$.
\begin{figure}[bt] 
	{\includegraphics*[width=3in]{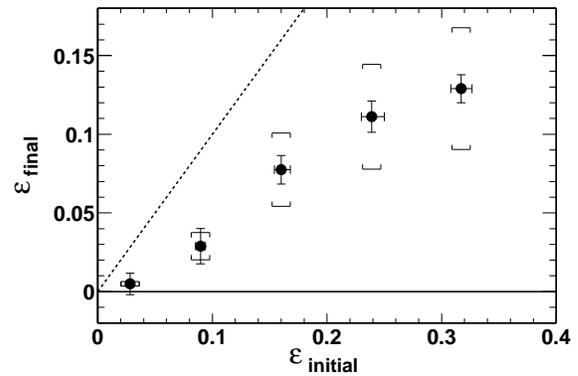}}
  \caption{Source eccentricity obtained with azimuthally-sensitive HBT ($\varepsilon_{\rm final}$) vs.~initial eccentricity
from a Glauber model ($\varepsilon_{\rm initial}$).  The most peripheral collisions correspond to the largest eccentricity. The dashed line indicates
$\varepsilon_{\rm initial}$ = $\varepsilon_{\rm final}$.  Uncertainties on the precise nature of space-momentum correlations lead to 30\% systematic errors on $\varepsilon_{\rm final}$ \cite{fabriceMike}.} \label{figure4}
\end{figure} 
Figure~\ref{figure4} shows the relation between the initial and final eccentricities obtained by averaging the three $k_T$ bins in Fig.~\ref{figure3}. The initial and final eccentricities 
exhibit a monotonic relationship, with more peripheral collisions showing a larger final 
anisotropy.   
Within this model-dependent picture, the source at freeze-out still retains some of its initial 
shape, indicating that the outward pressure and/or expansion time was not sufficient to quench 
the initial spatial anisotropy.  The large elliptic flow and small HBT radii observed at RHIC 
energies might favor a large pressure build-up in a short-lived system.  Also, out-of-plane 
freeze-out shapes tend to disfavor a long-lived hadronic rescattering phase following hydrodynamic expansion~\cite{teaney}.

In conclusion, we have performed an analysis of two-pion HBT interferometry relative 
to the reaction plane in Au+Au collisions at $\sqrt{s_{NN}}=200$ GeV.  The relative 
amplitudes of the HBT radius oscillation is largest for peripheral collisions, indicating
larger out-of-plane anisotropy in the pion source at freeze-out, for collisions with larger
initial spatial anisotropy.
No strong $k_T$ dependence of the relative oscillation amplitudes is observed.  The out-of-plane freeze-out shape of the source
indicates that the build-up of pressure and the evolution time of the 
expanding system are not sufficient to quench the initial geometry of the collision.  This 
information, taken together with the size of the source and anisotropies in momentum space, 
places significant constraints on future theoretical 
efforts to describe the nature and timescale of the collision's evolution.

We thank Drs.~U.~Heinz, P.~Kolb and U.~Wiedemann for enlightening discussions, and we thank the RHIC 
Operations Group and RCF at BNL, and the NERSC Center at LBNL for their support. This work was supported 
in part by the HENP 
Divisions of the Office of Science of the U.S.~DOE; the 
U.S.~NSF; the BMBF of Germany; IN2P3, RA, RPL, 
and EMN of France; EPSRC of the United Kingdom; 
FAPESP of Brazil; the Russian Ministry of Science and 
Technology; the Ministry of Education and the NNSFC 
of China; SFOM of the Czech Republic; DAE, DST, and 
CSIR of the Government of India; the Swiss NSF.


\begin{thebibliography}{99}
\bibitem{generalForm}  The general form of Eq.~\ref{eq:prattBertsch} contains
                       cross terms $R^2_{ol}$ and $R^2_{sl}$; however, without knowledge
                       of the $1^{\rm{st}}$-order reaction plane, these terms vanish by                          symmetry considerations~\cite{HHLW}.  $R^2_{os}$ is related to the
                       tilt angle between the emission direction and ``out" direction in
                       the Pratt-Bertsch decomposition of relative pair momentum \cite{heinzKolbPlb}.
\end{thebibliography}
\end{document}